\documentclass[pre,aps,twocolumn,showpacs,floatfix,10pt]{revtex4-1}

\usepackage{graphicx}
\usepackage{tensor}
\usepackage{amsmath, amsthm, amssymb}

\newcommand{\be}{\begin{equation}}
\newcommand{\ee}{\end{equation}}
\newcommand{\bea}{\begin{eqnarray}}
\newcommand{\eea}{\end{eqnarray}}

\begin{document}

\title{ Bethe approximation for a system of hard
rigid rods: the random locally tree-like layered lattice}  
\author{Deepak Dhar}
\affiliation{Department of Theoretical Physics, Tata Institute of Fundamental
Research, Homi Bhabha Road, Mumbai 400005, India}
\author{R. Rajesh}
\affiliation{The Institute of Mathematical Sciences, C.I.T. Campus,
Taramani, Chennai 600113, India}
\author{J\"{u}rgen F. Stilck}
\affiliation{Instituto de F\'{i}sica and National Institute of Science and
Technology for Complex Systems, Universidade Federal Fluminense, Av.
Litor\^anea s/n, 24210-346 Niter\'oi, RJ, Brazil}

\date{\today}

\begin{abstract}
We study the Bethe approximation for a system of long rigid rods of 
fixed length $k$, with only excluded volume interaction. For large enough
$k$, this system undergoes an isotropic-nematic phase transition as a 
function of density of the rods. The Bethe lattice, which is 
conventionally used to derive the self-consistent equations in the Bethe 
approximation, is not suitable for studying the hard-rods system, as it 
does not allow a dense packing of rods.  We define a new lattice, called 
the random locally tree-like layered lattice, which allows a dense 
packing of rods, and for which the approximation is exact. We find that 
for a $4$-coordinated lattice, $k$-mers with $k \geq 4$ undergo a 
continuous phase transition. For even coordination number $q \geq 6$, the 
transition exists only for $k \geq k_{min}(q)$, and is first order.
\end{abstract}


\maketitle

\section{\label{sec:intro}Introduction}

The study of the ordering transition in systems with only excluded 
volume interactions has a long history. Depending on the shape of the 
molecules involved, these systems can exhibit different ordered states 
and undergo transitions between them.  Many different shapes have been 
studied in literature: e.g., hard spheres \cite{mccoy2005}, hard squares 
\cite{pearce1988,fernandes2007}, hexagons \cite{baxter1980}, triangles 
\cite{nienhuis1999}, tetrominoes \cite{barnes2009}, rods \cite{vroege1992}, 
banana-shaped 
molecules \cite{maeda2005}, etc.  
A system of thin rigid cylindrical molecules in solution was shown
by Onsager to undergo a first order phase transition from an isotropic to a
nematic phase at sufficiently high densities, in the limit of large
aspect ratio \cite{onsager1949}.
Flory studied the hard rods problem on 
a lattice \cite{flory1956a}, still allowing the rods to have continuous 
orientations, and using a mean-field approximation again found a 
isotropic-nematic transition. Zwanzig studied a system of hard rods in 
the continuum, but restricting their orientations to a discrete set 
\cite{zwanzig1963}, also finding a discontinuous transition to a nematic 
phase as the density of rods is increased. The relation between
continuous and discrete models is discussed in
Ref.~\cite{shundyac2004}.

For continuous models, in dimensions $d \geq 3$, if the aspect ratio of 
the rods is high enough, the existence of an isotropic--nematic
transition is well accepted \cite{vroege1992}.
In two dimensions, in a system of hard needles in a plane, the Mermin--Wagner
theorem \cite{mermin1966} disallows long range orientational order, but at
sufficiently high densities, the system shows a Kosterlitz-Thouless type
ordering transition, and the orientational correlations decay with
distance as a power law \cite{frenkel1985,khandkar2005}.
If the rigid rods are placed on a lattice, then only a discrete set of
orientations is possible, and the existence of a long-range ordered phase
is not disallowed by the Mermin--Wagner theorem.
Rods occupying $k$ consecutive sites along any one lattice direction
will be called $k$-mers.
For dimers ($k=2$), it can be shown rigorously that orientational 
correlations decay exponentially, except in the limit when all sites are 
occupied by rods.  In the case of full coverage, the orientational 
correlations have a power law tail in all dimensions $ d \geq 2$ 
\cite{lieb1972,huse2003}.

The question whether the lattice model of rigid rods with $k \gg 2$ 
undergoes a phase transition remained unsettled for a long time
\cite{degennesBook}. 
Clearly, the number of ways of covering the planar square lattice fully 
with $k$-mers grows exponentially with the size of the system, and most 
of these configurations are not ordered. Thus, at the maximum packing 
density, which was presumed to favour ordering most, the system shows no 
ordering. Recently, Ghosh and Dhar argued that $k$-mers on a square 
lattice, for $k \geq 7$, would undergo two phase transitions, and 
the nematic phase would exist for only an intermediate range of densities 
$\rho_1^* < \rho < \rho_2^*$, where $\rho_1^* \sim 1/k$
\cite{ghosh2007}.
Subsequently, Fernandez et. al. 
\cite{fernandez2008a, fernandez2008b, fernandez2008c,linares2008}, through
extensive simulations and approximate methods,  showed that
the square lattice transition is Ising like,  or 
equivalently in the liquid-gas universality class \cite{fischer2009}. On 
the triangular and hexagonal lattices, there are three degenerate fully 
aligned ground states, and the transition was shown to be in the $q=3$ 
Potts model universality class \cite{fernandez2008a,fernandez2008b}. 
The existence of a transition to a state with orientational order was shown
rigorously for a polydispersed hard rod system on a two dimensional square
lattice \cite{velenik2006}.
The second transition from nematic to 
unordered state at $\rho_2^*$ is much less understood. In Ref.~\cite{ghosh2007},
a variational estimate of the entropy of the nematic 
and ordered states suggests that $1 -\rho_2^*$ should vary as $1/k^2$ 
for large $k$, but not much is known about the nature of transition. 
Linares et. al. estimated that $ 0.87 \leq \rho_2^* \leq 0.93$ for $k 
= 7$, and proposed an approximate functional form for the entropy as a 
function of the density \cite{linares2008}.

Given the paucity of exact results about the second transition, it seems 
worthwhile to investigate this problem using well-known approximation 
methods, like the Bethe approximation (BA).  However, we realized that 
there are problems in the application of the conventional BA techniques, 
even to the first isotropic--nematic transition. In particular, 
packing $k$-mers on the Bethe lattice with a density close to one seems 
difficult, and the nature of the high-density phase on the Bethe lattice 
is not very clear.  Another way to study BA is to work with random 
graphs of fixed-coordination number
\cite{baillie1994,dhar1997,dembo2010a,dembo2010b}. 
In this case, there is 
no surface, but in this case as well, one does not get full coverage in the 
limit when the activity of $k$-mers tends to infinity. This has led us to 
introduce a new lattice, to be called the random locally tree-like 
layered lattice (RLTL lattice). We study $k$-mers on the RLTL and show that
there is no difficulty in defining the high-density phase for the 
$k$-mers and that the BA is exact on this lattice. 
In fact, we find many ways to  cover the lattice with k-mers. There is a
finite entropy per site in the fully packed limit, and in a typical fully
packed configuration, all k-mers are not all in the same orientation.

The rest of the paper is organised as follows. We 
begin with a recapitulation of 
the conventional approaches to the Bethe approximation, and the Bethe 
lattice in Sec.~\ref{bethelat}, and derive self-consistent equations between 
different correlation functions. In Sec.~\ref{rltlconstruction}, 
we describe the RLTL 
lattice where every site in the lattice has the same even coordination number 
$q$. In Sec.~\ref{randomlat}, we derive the exact annealed 
partition function of the 
problem on this lattice, and show that the resulting self-consistent 
equations are the same as obtained earlier for the Bethe lattice. In 
Sec.~\ref{intrans}, we analyse the behaviour of these equations for the 
RLTL with 
coordination number $4$, and show that it undergoes a continuous 
transition for $k>4$. In Sec.~\ref{qhigherfour}, we discuss the case 
$q \geq 6$. We find 
that in this case, the system undergoes a first order transition for $k 
> k_{min}(q)$. Section~\ref{discussion} contains some concluding remarks.

\section{$k$-mers on a $4$-coordinated Bethe lattice}
\label{bethelat}

The BA was initially proposed as an ad-hoc decoupling approximation (the 
pair approximation) \cite{bethe1935,rushbrooke1938}.  It was realized that the 
approximation becomes exact for the Bethe lattice, which is the part of 
a uniformly branching tree graph far away from the surface
\cite{kurata1953,rushbrooke1955}. 
This 
is important, as this ensures that the approximation can not give rise 
to unphysical consequences, like non-positive specific heat. Also, there 
is a consistent scheme, and two different answers for same quantity 
cannot be obtained. Given that in a uniformly branching tree, most of 
the sites are a finite distance from the surface, disentangling the 
surface contribution from that of sites deep inside requires some care
\cite{runnels,eggarter,zittartz}.
There are some 
prescriptions for separating the surface contribution from bulk given in 
literature \cite{gujrati1995}. A rather careful and detailed discussion for 
the case of the Ising model on the Bethe lattice is given in Chapter~4 of
Ref.~\cite{baxterBook}.
One considers correlation 
functions deep inside a Cayley tree, whose values are independent of the 
boundary conditions on the surface of the tree. This can be realized for 
all non-zero external magnetic fields. Then thermodynamical quantities 
like the free energy are obtained by integrating these correlation 
functions with respect to suitable conjugate field.

A more recent approach to BA is to treat it as the free energy for a 
random graph 
on $N$ sites, where each site in the graph has the same 
degree. One can show that in the limit $N$ tends to infinity, for a 
randomly picked site in the graph, the length of the shortest loop 
increases as $\log N$. Then, the graph looks like a loop-less of uniform 
coordination number, up to a distance of order $\log N$.  In the limit 
of large $N$, it looks like a Bethe lattice, except that there is no 
surface. This makes study of the different models on the Bethe lattice 
numerically feasible \cite{baillie1994,dhar1997,dembo2010a,dembo2010b}.

\begin{figure} 
\includegraphics[width=\columnwidth]{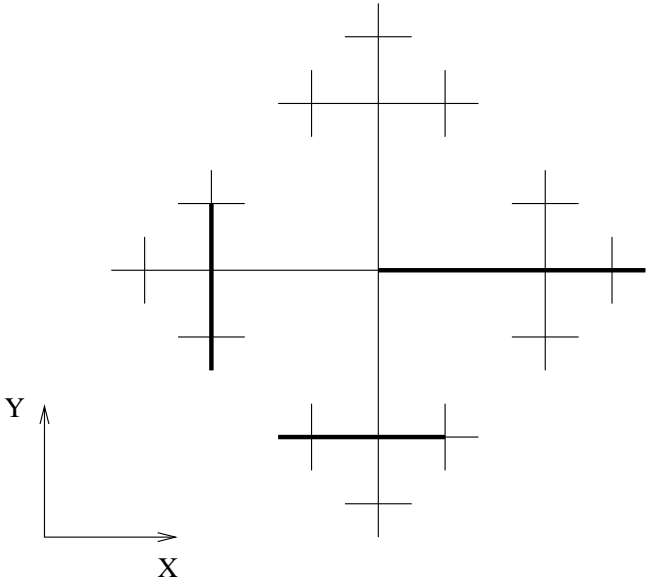} 
\caption{A 
four-coordinated Cayley tree with four generations is shown. Two X-mers 
(horizontal thick solid lines) and one Y-mer (vertical thick solid line) 
with $k=4$ are placed on the tree. The weight of the configuration is 
$z_1^2 z_2$.} 
\label{bl}
\end{figure}

We first derive the self consistent equations satisfied by densities for 
the model of $k$-mers on a Bethe lattice of coordination number four in 
the conventional way. Generalisation to higher even coordination number 
is straightforward. A Bethe lattice corresponds to the core of the 
Cayley tree with the same coordination number. The sites of the Cayley 
tree may be ordered by its generation, starting with $m=0$ for the sites 
on the surface and ending at $m=M$ for the central site of a tree with 
$M$ generations. Each site of the Cayley tree, other than those on the 
surface, has $4$ bonds attached to it. Two of them will be said to be of 
type X and the remaining two to be of type Y.  A $k$-mer occupies $k-1$ 
consecutive bonds ( and $k$ sites) of the same type. We associate an 
activity $z_1$ with an X-type $k$-mer (referred to as X-mer in the 
following) and an activity $z_2$ with a Y-type $k$-mer (Y-mer). No two 
$k$-mers may overlap, so that each site of the lattice has at most one 
$k$-mer passing through it. Fig.~\ref{bl} shows a four-coordinated Cayley 
tree occupied by three 4-mers.

The Cayley tree is composed of four rooted subtrees which are connected 
to the central site. We define partial partition functions for these 
rooted subtrees with fixed configuration of the root bond.  We specify 
the type of the root bond and the configuration of the root bond. The 
root bond may be empty, or if it is occupied by a $k$-mer, then we specify 
the number of sites in earlier generations already occupied by this 
$k$-mer. There are $2 k$ such partial partition functions, which we denote 
by $g_{x,j}$ and $g_{y,j}$, where the first subscript denotes the type 
of the root bond and $j=1,\ldots,k-1$ is the number of sites that are 
currently part of the $k$-mer.  $j=0$ corresponds to the case when the 
root bond is empty.

\begin{figure}
\includegraphics[width=7cm]{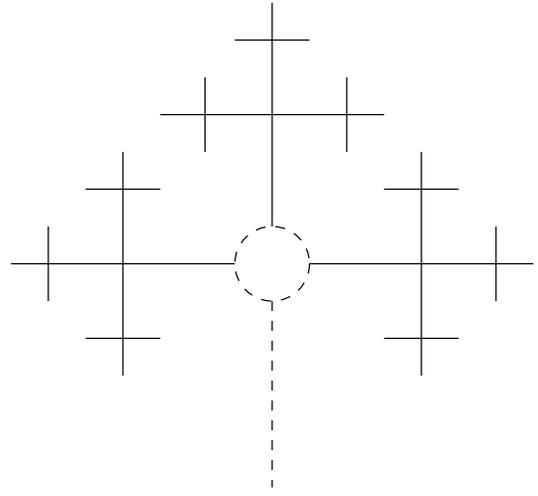}
\caption{Building a subtree with $M+1=4$ generations of sites by connecting 
$q-1=3$ subtrees with $M=3$ generations of sites to a new root site and 
bond (denoted by dashed lines). If the root bond of the new subtree is in 
the $Y$-direction, then 
two of the existing subtrees will have their root bond in the 
$X$-direction.}
\label{rr}
\end{figure}

\begin{figure}
\includegraphics[width=\columnwidth]{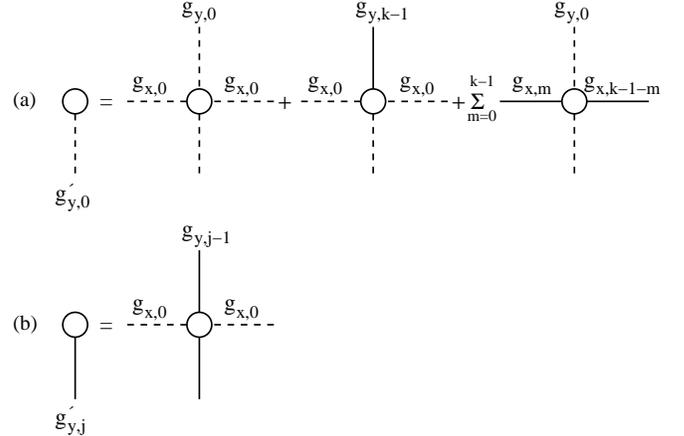}
\caption{Diagrammatic representation of Eq.~(\ref{eq:grecursion}) 
for $g^{\prime}_{y,j}$, $j=0,\ldots,k-1$.
} 
\label{fig:recursion1}
\end{figure}

The recursion relations obeyed by the partial partition functions are 
obtained by building a subtree with one additional generation by 
connecting 3 subtrees to a new root bond and new root site. This process 
is illustrated in Fig.~\ref{rr}. We attach a weight $z_1^{1/k}$ and 
$z_2^{1/k}$ with each occupied site of and X-mer and Y-mer respectively. 
The recursion relations are (see Fig.~\ref{fig:recursion1})
\begin{subequations}
\label{eq:grecursion}
\begin{eqnarray}
g'_{x,0}&=&(g_{x,0}+ z_1^{1/k}g_{x,k-1})g_{y,0}^2 \nonumber \\
&+& z_2^{1/k} g_{x,0}\sum_{j=0}^{k-1}g_{y,j}g_{y,k-1-j}, \\
g'_{y,0}&=&(g_{y,0}+z_2^{1/k} g_{y,k-1})g_{x,0}^2 \nonumber \\
&+& z_1^{1/k} g_{y,0} \sum_{j=0}^{k-1}g_{x,j}g_{x,k-1-j}, \\
g'_{x,j}&=&z_1^{1/k} g_{x,j-1}g_{y,0}^2, \quad j=1,\ldots,k-1, \\
g'_{y,j}&=&z_2^{1/k} g_{y,j-1}g_{x,0}^2, \quad j=1,\ldots,k-1,
\end{eqnarray}
\end{subequations}
where the prime denotes partial partition functions 
of subtrees with one additional generation of sites. 

The partial partition functions are multiplied by the appropriate 
activity each time a $k$-mer grows such that the weight of a $k$-mer is 
$z_1$ or $z_2$ depending on its type.  The partial partition functions 
grow exponentially with the number of iterations. To calculate the 
densities in the core of the tree (Bethe lattice), it is useful to 
define ratios of partial partition functions
\be
R_{i,j}=\frac{g_{i,j}}{g_{i,0}}, ~i=x,y~\mathrm{and}~j=0,\ldots,k-1,
\ee
such that $R_{x,0}=R_{y,0}=1$. The recursion
relations obeyed by the ratios are easily derived from
Eq.~(\ref{eq:grecursion}):
\begin{subequations}
\label{eq:recursion_r}
\begin{eqnarray}
R'_{x,j}&=&\frac{z_1^{1/k} R_{x,j-1}}{D_x}, \quad j=1,\ldots,k-1,\\
R'_{y,j}&=&\frac{z_2^{1/k} R_{y,j-1}}{D_y}, \quad j=1,\ldots,k-1,
\end{eqnarray}
\label{rrel}
\end{subequations}
where 
\begin{subequations}
\begin{eqnarray}
D_x&=&1+z_1^{1/k} R_{x,k-1}+ z_2^{1/k} \sum_{j=0}^{k-1}R_{y,j}R_{y,k-1-j},\\
D_y&=&1+z_2^{1/k} R_{y,k-1}+ z_1^{1/k} \sum_{j=0}^{k-1}R_{x,j}R_{x,k-1-j}.
\end{eqnarray}
\end{subequations}

We study the fixed points of these recursion relations. The fixed point 
corresponds to distance of the surface sites very large, or equivalently 
to sites in the deep interior of the Cayley tree. Also, if only one 
single stable fixed point exists, the boundary conditions at surface 
sites (which determine the initial values of the recursions), have no 
effect on the behaviour at points deep inside \cite{eggarter,zittartz}.

A natural choice of the starting weights for the recursion equations are 
such that all $k$-mers are completely contained within the Cayley tree. 
This corresponds to $g_{x,0} = g_{y,0} = 1$, $g_{x,1} = z_1^{1/k}$, and 
$g_{y,1} = z_2^{1/k}$, with $g_{x,j}=g_{y,j}= 0$ for all $j>1$. However, 
for these choices of initial conditions, the recursion relations under 
iteration converge to a fixed point that corresponds to the isotropic 
system because the initial conditions themselves are symmetric. The
isotropic fixed point is stable with respect to perturbations where
the initial conditions are still symmetric with respect to the two
directions. However, it is unstable to asymmetric perturbations.

For generic initial conditions, we find that on repeatedly iterating the
recursion relations, the R's converge to a stable fixed point $R^*$, which
may be determined by iterating 
Eq.~(\ref{eq:recursion_r}). Then,
\begin{subequations}
\bea
R_{x,j}^* &=& \left(\frac{z_1^{1/k}}{D_x}\right)^j \equiv \alpha_x^j,
~j=0,\ldots, k-1,\\
R_{y,j}^* &=& \left(\frac{z_2^{1/k}}{D_y}\right)^j \equiv \alpha_y^j,
~j=0,\ldots, k-1,
\eea
\end{subequations}
where the variables $\alpha_x$, $\alpha_y$ satisfy the 
equations 
\begin{subequations}
\begin{eqnarray}
\alpha_x[1+z_1^{1/k}\alpha_x^{k-1}+k z_2^{1/k}\alpha_y^{k-1}]&=&z_1^{1/k}, \\
\alpha_y[1+z_2^{1/k}\alpha_y^{k-1}+kz_1^{1/k}\alpha_x^{k-1}]&=&z_2^{1/k}.
\end{eqnarray}
\label{fpe}
\end{subequations}

Knowing the fixed point solution, the density at the central site may be
calculated. The grand canonical partition function of the system on the
Cayley tree is given by
\begin{eqnarray}
\Xi&=& 
g_{x,0}^2 g_{y,0}^2+
z_1^{1/k} g_{y,0}^2  \sum_{j=0}^{k-1}g_{x,j}g_{x,k-1-j} 
\nonumber\\
&+&
z_2^{1/k} g_{x,0}^2 \sum_{j=0}^{k-1}g_{y,j}g_{y,k-1-j}.
\label{gcpfct}
\end{eqnarray}
Then, the densities $\rho_x$ ($\rho_y$) of sites that are part of
X-mers (Y-mers) are given by
\begin{subequations}
\label{eq:bethe_density}
\bea
\rho_x&=& \frac
{z_1^{1/k} g_{y,0}^2 \sum_{j=0}^{k-1}g_{x,j}g_{x,k-1-j}} {\Xi},\\
\rho_y&=& \frac
{z_2^{1/k} g_{x,0}^2 \sum_{j=0}^{k-1}g_{y,j}g_{y,k-1-j}} {\Xi},
\eea
\end{subequations}

At the fixed points, the densities simplify to
\begin{subequations}
\begin{eqnarray}
\rho_x&=&\frac{k z_1^{1/k}\alpha_x^{k-1}}{1+kz_1^{1/k}\alpha_x^{k-1}+
k z_2^{1/k}\alpha_y^{k-1}}, 
\\ 
\rho_y&=&\frac{kz_2^{1/k}\alpha_y^{k-1}}{1+k z_1^{1/k}\alpha_x^{k-1}+
k z_2^{1/k}\alpha_y^{k-1}}.
\end{eqnarray}
\label{rho12}
\end{subequations}

Eliminating $\alpha_x$, $\alpha_y$ from Eqs.~(\ref{rho12}) and 
Eqs.~(\ref{fpe}), we obtain
\begin{subequations}
\begin{eqnarray}
z_1(1-\rho_x-\rho_y)^k&=&\frac{\rho_x}{k}\left(1-\frac{k-1}{k}\rho_x
\right)^{k-1}, \\
z_2(1-\rho_x-\rho_y)^k&=&\frac{\rho_y}{k}\left(1-\frac{k-1}{k}\rho_y
\right)^{k-1}.
\end{eqnarray}
\label{eq:bethe}
\end{subequations}

The Bethe lattice solution described above is not very satisfactory.  In 
particular, in the limit of large $z_1$ and $z_2$, one gets the fraction 
of sites occupied by $k$-mers tends to $1$. However, it is easy to see 
that this can only be achieved with a very special choice of boundary 
conditions at the surface of the tree. 

Also, if we consider a tree with coordination number $q \geq 6$, in the 
large activity limit, then  at any root vertex of a subtree, the process of 
exchanging different branches is clearly a symmetry operation on the 
lattice.  It is easy to see that the presence of this local symmetry 
implies that there can be no nematic order in the deep inside region, 
with one type of bonds preferentially occupied.  This makes this 
approach unsuitable for the studies of the hard rod problem.

\section{\label{rltlconstruction}The RLTL lattice construction}

\begin{figure}
\includegraphics[width=\columnwidth]{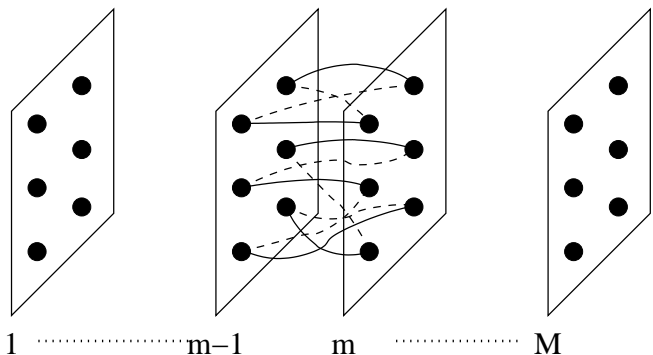}
\caption{Schematic diagram of the random lattice with $N=6$ sites per
layer and coordination number $q=4$. A possible configuration of
bonds between layers $m-1$ and $m$ is shown, with the solid and dotted lines 
being bonds of two different types.} 
\label{fig01}
\end{figure}

For simplicity, we discuss the random lattice with coordination number 
$q=4$. Generalisation to other coordination numbers is straight forward. 
We consider a set of $M$ layers, numbered from $1$ to $M$, with $N$ 
sites in each layer. A layer $m$ is connected to the adjacent layer 
$m-1$ by $N$ bonds of type $X$ and $N$ bonds of type $Y$. The 
connections are made by randomly pairing each site of the mth layers 
with exactly one site in the $(m-1)$-th layer with an X-bond, similarly 
randomly pairing each site with a site in the neighbouring layer using 
Y-bonds. The total number of such possible pairings is $(N!)^2$. This is 
illustrated in Fig.~\ref{fig01}.

We can impose open or periodic boundary conditions. For a $2q$ 
coordinated lattice, there are $(N!)^{q M}$ different possible graphs, 
in the case of periodic boundary conditions (layer $M$ is connected to 
layer $1$), and $(N!)^{q (M-1)}$ different possible graphs for the case of 
open boundary conditions. We consider annealed models on this lattice, 
thus we average the partition function over all possible configurations 
of the bonds.

We can associate different degrees of freedom with the vertices, and 
consider statistical mechanical models on these lattices.  For example, 
we can attach an Ising spin $S_j$ to each vertex $j$ of the lattice, and 
define the Hamiltonian to be $H = - J \sum_{nn} S_i S_j$, where the sum 
is over all nearest neighbour pairs. For a particular realization ${\cal 
R}$ of bonds, let the partition function be ${\cal Z}_{\cal R}(M,N)$. 
We average the partition function over all possible configurations of bonds.
As we are averaging the partition function, and not 
the logarithm of the partition 
function, this is similar to the annealed average over different  
bond configurations. Thus,
\begin{equation}
Z_{av}(M,N) = \frac{1}{N_{{\cal R}}} \sum_{{\cal R}} {\cal Z}_{\cal
R}(M,N),
\end{equation}
where $N_{{\cal R}}$ is the number of different bond configurations on 
the lattice.

To take the thermodynamic limit, we let $M$ and $N$ tend to infinity. 
The mean free energy per site $f$ is defined as the limit
\begin{equation}
f = - kT \lim_{M,N \rightarrow \infty} \frac{1}{MN} \log Z_{av}(M,N).
\end{equation}

We note that each site, say in layer $m$, has exactly one X-bond, and 
one Y-bond connecting it to sites in the layer $m+1$.  It can happen 
that these bonds fall on the same site. However, the fractional number 
of such cases is only $1/N$, and thus, the expected number of loops of  size 
$2$ on this lattice is M. For $M$ even, the graph is bipartite, and 
there are no loops of odd perimeter.  It is easy to verify that the 
expected number of loops of perimeter $4$ on this lattice is $5 M$. In 
general, the number of loops of perimeter $\ell$ per site of the lattice 
varies as $\lambda^{\ell}/N$, where $\lambda$ is the self-avoiding walk 
growth constant \cite{bollobasBook}.

We, thus, see that for large $N$, there are very few short loops in the 
lattice. For a randomly picked site, the size of the shortest loop going 
through that site is of order $\log N$, and this goes to infinity, as 
$N$ goes to infinity. Since there are very few short loops, the structure 
of the lattice locally is that of a regular branching tree, and it 
locally looks like the Bethe lattice.

The correlation functions can be defined as usual. Consider the example 
of the Ising model defined above. We consider two sites $i$ and $j$, and 
consider the two-point correlation function $\langle S_i S_j \rangle$. 
Since this correlation function has to be averaged over all assignments 
of bonds, it can depend only on the difference in layer numbers 
of the sites. In particular, it has the same value for all sites $i$ and 
$j$ in the same layer. Thus, the expected factorisation property of the 
correlation functions, which is the essence of the Bethe approximation, 
is built into the definition of the lattice.

\section{$k$-mers on the RLTL lattice with $q =4$}
\label{randomlat}

We consider a $k$-mer model on the RLTL lattice. A $k$-mer occupies 
$(k-1)$ consecutive bonds of the same type. As earlier, we 
associate an activity $z_1$ with an X-mer and 
an activity $z_2$ with a Y-mer.

Let $x_m$ be the number of X-mers with topmost ($1^{st}$) site in the 
$m$-th layer, and $y_m$ is number of Y-mers with topmost site in the 
$m$-th layer. We will denote by $X_m$ and $Y_m$, the number of sites in 
the $m$th layer, occupied by X-type, and Y-type bonds respectively, but 
where the site is not the topmost site of the $k$-mer.  Clearly, we have
\begin{equation}
X_m= \sum_{j=1}^{k-1} x_{m-j}, ~~~Y_m= \sum_{j=1}^{k-1} y_{m-j}.
\end{equation}

We can adopt the convention that $x_m= y_m =0$, for $m \leq 0$. Then the 
above equation holds for all $m, 1 \leq m \leq M$. Note that if we want 
the $k$-mers to be fully contained in the lattice, we must also have 
$x_m= y_m =0$, for $m \geq M-k+2$.

To calculate the partition function, consider the operation of adding 
an additional layer. We, thus, specify the full set of $2M$ values 
$\{x_m,y_m\}$ for all $m$. The total statistical weight of 
configurations that contribute to a particular set $\{x_m,y_m\}$ will be 
denoted by $C(\{x_m,y_m\})$.

We calculate $C(\{x_m,y_m\})$ recursively.  Let us imagine that we have 
constructed the configuration of the lattice to layer $m-1$, and now 
we add the layer $m$. We sum over the configurations of bonds, and
that of X-mers and Y-mers on these bonds at the same time.

(i)
We note that in the $(m-1)$-th layer, there are $X_m$ sites that are
occupied by X-mers which protrude to layer $m$. The first of these
sites is connected to a randomly picked site in the lower layer in $N$
ways using an X-bond.  Then this site in the lower layer is occupied
by the extension of that X-mer. The next can be connected to one of
the unoccupied  sites   in the lower layers in only $(N-1)$
ways. Similarly the third, and so on. The number of ways to do this is: 
\[
\frac{N!}{(N-X_m)!}.
\]

(ii)
Now we take the $Y_m$ Y-mers in the $(m-1)$-th layer that are extending 
down to the lower layer. Number of ways of extending these to the 
$(N-X_m)$ unoccupied sites below is clearly
\[
\frac{(N-X_m)!}{(N-X_m -Y_m)!}.
\]

(iii) 
Connect the remaining $N-X_m$ $X$-bonds between layers $m-1$ and $m$
to sites in layer $m$ not yet connected by $X$ bonds. The number of
ways to do this is 
\[
(N-X_m)!.
\]

(iv)
Repeat the last procedure with the $N-Y_m$ remaining $Y$-bonds between
the layers $m$ and $m+1$. The number of 
ways to do this is 
\[
(N-Y_m)!.
\]

(v)
Finally, the $(N-X_m -Y_m)$ sites in layer $m$, that are unoccupied so 
far, are divided into three groups: $x_{m}$ topmost sites of new 
X-mers, $y_{m}$ topmost sites of new Y-mers, and the unoccupied 
sites. Clearly the number of ways to do this is:
\[
\frac{(N- X_m -Y_m)!}{x_{m}!y_{m}!(N- X_m- Y_m-x_{m}-y_{m})!}.
\]

The product of these factors gives the total number of ways of adding
the $m$-th layer as: 
\begin{equation}
\frac{N!(N-X_m)!(N-Y_m)!}{x_{m}!y_{m}!  (N-X_m- Y_m-x_{m}-y_{m})!}.
\end{equation}
Finally, multiplying these factors for different $m$, we get the
configurations with specified $\{x_m,y_m\}$ have total number of
configurations   $C(\{x_m,y_m\})$ given by  
\begin{equation}
C(\{x_m,y_m\})= \! \prod_{m=1}^M \frac{N!(N-X_m)!(N-Y_m)! }
{x_{m}! y_{m}! (N- \! X_m- \! Y_m-x_{m}-y_{m})!}.
\label{eq:C}
\end{equation}
Putting in the corresponding activity factors, the grand partition
function for the whole lattice with $M$ layers  is 
\begin{equation}
Z_{av} = \frac{1}{(N!)^{2M-2}} \sum_{\{x_m,y_m\}} C(\{x_m,y_m\})
z_1^{\sum x_m} z_2^{\sum y_m}, 
\label{zrl}
\end{equation}
where the sum is over all possible values of $\{x_m,y_m\})$, 
and we have divided the multiplicity factor by the number of configurations of
the random lattice  $(N!)^{2M-2}$, to get the average partition function. 
Note that  $Z_{av}=1$ for $z_1 = z_2 =0$, as expected.

The summation over $\{x_m,y_m\}$ yields at most a factor of order $N^{2 
M}$. Since the summand is of order $\exp(NM)$, for large $N$, we can 
ignore the summation over $\{x_m,y_m\}$, and replace the summation with 
the largest term, with negligible error. For the summand to be 
maximum with respect to $x_j$, we set
\begin{equation}
\frac{C(\{x_m + \delta_{m,j} , y_m\}) z_1}{ C(\{x_m , y_m\})} \approx 1.
\end{equation}

This gives
\begin{equation}
z_1  \prod_{s=0}^{k-1}  \frac {(N - X_{j+s}-x_{j+s} -Y_{j+s} -y_{j+s})}
{  (N- X_{j+s})}
= \frac {x_j +1}{(N-X_j)}, 
\label{eq:7}
\end{equation}
and similarly
\begin{equation}
z_2 {\prod_{s=0}^{k-1}} \frac{
(N - X_{j+s}-x_{j+s} -Y_{j+s} -y_{j+s})}
{ (N- Y_{j+s})}
= \frac{(y_j +1)}{(N-Y_j)}. 
\label{eq:8}
\end{equation}

Writing the maximising values as $f^*_j= x_j /N $ and $g^*_j= y_j/N$, the 
equations satisfied by $f^*_j$ and $g^*_j$ in the limit $N$ tends to 
infinity are
\bea
\label{eq:7a}
z_1 \prod_{s=0}^{k-1}   \frac
{1 -\rho_x(j+s)  - \rho_y(j+s)}{ 1- \rho_x(j+s)+ f^*_{j+s}}
&= & 
\frac{f^*_j}{1 -\rho_x(j)+ f^*_j }, \\
\label{eq:8a}
z_2 \prod_{s=0}^{k-1}   {\frac
{1 -\rho_x(j+s)  - \rho_y(j+s)}{ 1- \rho_y(j+s)+ g^*_{j+s}}}
&= & 
\frac {g^*_j}{1 - \rho_y(j) + g^*_j}, 
\eea
where $\rho_x(m)$  and $\rho_y(m)$ are  the fractions of sites in
layer $m$ covered by X-mers and Y-mers respectively. Clearly,
\begin{equation}
\rho_x(j) = \sum_{s=0}^{k-1} f^*_{j-s} ,~~\rho_y(j) = \sum_{s=0}^{k-1} 
g^*_{j-s}.
\end{equation}

These equations connect $f^*_j$ and $g^*_j$ to their value of 
$f^*_{j+s}$ and $g^*_{j+s}$, with $s = 1$ to $k-1$. These may be 
considered as recursion equations for $f^*_j, g^*_j$. These recursions 
work in the direction of decreasing $j$.

These equations have a simple interpretation. In the equilibrium state, 
$z_1$ is the ratio of the probability
that a randomly chosen site will be the head of a 
X-mer to the probability that an X-mer can be placed with 
this site as the head. The probability that the chosen site is 
empty is $[ 1 -\rho_x(j) -\rho_y(j)]$.  Given that a given site is empty in 
layer $(j'-1)$, the conditional probability that the site connected to 
it in the layer $j'$ by an X-bond is also empty is $\frac{ 1 -\rho_x(j') 
-\rho_y(j')}{1 - \rho_x(j') + f^*_{j'}}$. Multiplying these 
probabilities for $(k-1)$ consecutive layers, we get the probability 
that a given site in layer $j$ can be the head of an X-mer to be
\begin{equation}
\nonumber
\frac{
\prod_{s=0}^{k-1} [1 - \rho_x(j+s) -\rho_y(j+s)]} {\prod_{s=1}^{k-1}
[ 1 - \rho_x(j+s) + f^*_{j+s})]}. 
\end{equation}
The probability that the chosen site in the $j$th layer is the head of 
an X-mer is $f^*_j$. The ratio of these is $1/z_1$, which gives
Eq.~(\ref{eq:7}).
  
The simplest solution of this is  a fixed point solution with $ f^*_j
= f^*$, $g^*_j= g^*$, independent of $j$ (away from the boundaries). 
Then $f^*, g^*$ satisfy the equations
\begin{subequations}
\begin{eqnarray}
z_1 ( 1 - k f^* - k g^*)^k &=& f^*  [ 1 - (k-1) f^*]^{k-1}, \\
z_2 ( 1 - k f^* - k g^*)^k &= & g^* [ 1 - (k-1) g^*]^{k-1}.
\end{eqnarray}
\label{eq:rhoq4}
\end{subequations}

These equations are the same as Eq.~(\ref{eq:bethe}), and we have 
recovered the Bethe approximation. However, for the RLTL lattice, the 
limit of fully packed lattice is well-defined, and causes no 
difficulties.

From Eq.~(\ref{eq:C}), the entropy per site (divided by $k_B$) is easily 
seen to be
\begin{eqnarray}
s(\rho_x,\rho_y)&=&\left(1-\frac{k-1}{k}\rho_x\right)\ln
\left(1-\frac{k-1}{k}\rho_x\right)
\nonumber \\
&+& \left(1-\frac{k-1}{k}\rho_y\right)\ln
\left(1-\frac{k-1}{k}\rho_y\right)
\nonumber \\
&-&  (1\!-\!\rho)\ln (1\!-\!\rho)
-\frac{\rho_x}{k}\ln
\frac{\rho_x}{k}-\frac{\rho_y}{k}\ln \frac{\rho_y}{k},
\label{entrop4}
\end{eqnarray}
where $\rho=\rho_x+\rho_y$ is the total density. The same expression for
entropy was obtained by DiMarzio \cite{dimarzio1961,fernandez2008b} 
who used an approximate counting technique for counting configurations 
on cubic lattices in any dimension. Also, Eq.~(\ref{entrop4})
coincides with the expression for entropy that one obtains by using
Gujrati's prescription for calculating free energies on the Bethe
lattice \cite{gujrati1995}.

It is easy to see that this expression for the entropy is not everywhere 
convex. When the value of $s(\rho_x,\rho_y)$, calculated as above, turns 
out to be in the non-convex region, it is easily seen that a much larger 
contribution to the partition function comes from $\{x_m,y_m\}$ that are 
not nearly uniform, and in a canonical ensemble at fixed $\rho_x$ and
$\rho_y$, the lattice will show phase separation, with one 
region having higher density than the other. The net effect of this to 
replace non-convex parts of the entropy function by a convex envelope 
construction. Thus, we write the true entropy $\tilde{s}(\rho_x,\rho_y)$ 
as
\begin{equation}
\tilde{s}(\rho_x,\rho_y) = {\cal CE}\left[ s(\rho_x,\rho_y)\right],
\end{equation}
where ${\cal CE}$ denotes convex envelope.

\section{Isotropic--Nematic Transition}
\label{intrans} 

In this section, we analyse the isotropic--nematic transition when the 
coordination number is four. From Eq.~(\ref{entrop4}), the expression 
for entropy, it is straight forward to determine the ordering that has 
maximum entropy.

\begin{figure}
\includegraphics[width=\columnwidth]{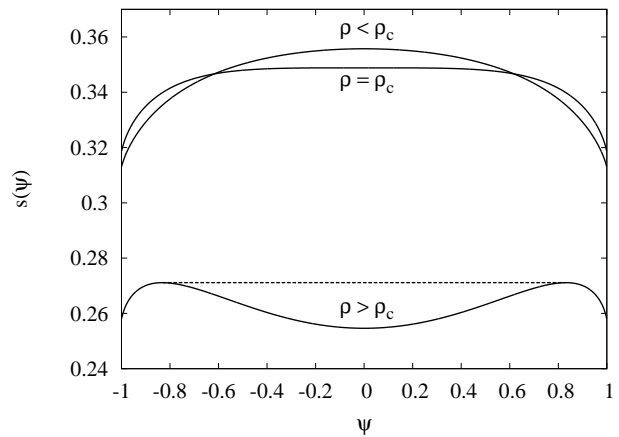}
\caption{Entropy as a as a function of the order parameter $\psi$ for
different densities. The data are for $q=4$, $k=4$ when the transition
is continuous. Entropy has one peak for small densities and two symmetric
peaks for large densities. The dotted lines denotes the convex envelope. } 
\label{fig:entropy_order}
\end{figure}

Consider the system at a fixed density $\rho =\rho_x + \rho_y$. We 
define the order parameter $\psi$ by
\begin{equation}
\psi = \frac{\rho_x - \rho_y}{\rho}.
\end{equation}
Then, it is easy to study the variation of $s(\rho_x, \rho_y)$ as a 
function of $\psi$, for fixed $\rho$. We find that for small $\rho$, 
the entropy has a single maximum at $\psi =0$, but for large enough 
$\rho > \rho_c$, it develops two symmetrically placed maxima (see
Fig.~\ref{fig:entropy_order}).
For small $\psi$, we can expand the entropy in a power series 
around $\psi =0$ as in the standard Landau treatment
\begin{equation}
s(\rho_x,\rho_y) = A(\rho) - \psi^2 B(\rho) + \psi^4 C(\rho) +\ldots.
\end{equation}

The coefficient of the quadratic term $B(\rho)$ changes sign at 
$\rho=\rho_c$, and is negative for larger $\rho$. Thus, for $\rho > 
\rho_c$, we have a nematic phase, with nonzero value of $\psi$. The 
critical exponent for $\beta$ for the order parameter takes the 
classical Landau theory value $1/2$.

The value of $\rho_c$ can be determined easily from Eq.~(\ref{eq:rhoq4}). 
For $z_1 =z_2 = z$, we note that both $f^*$ and $g^*$ 
are solutions to an equation of the form
\begin{equation}
x \left( 1 - \frac{k-1}{k} x \right)^{k-1} = constant.
\end{equation}

The left hand side of the equation is a function of $x$ that starts 
at $0$ for $x=0$, increases to a maximum value, and then decreases 
monotonically, and reaches a positive finite value $(1/k)^{k-1}$, for 
$x=1$. When the right hand constant is small enough, there is only one 
real valued solution of this equation, and $f^* = g^*$. For a range of 
values of the constant, there are exactly two distinct solutions. At the 
critical point, the two solutions are degenerate. This occurs where the 
function is maximum, i.e., at
\begin{equation}
x^* = \frac{1}{k-1}.
\end{equation}

\begin{figure}
\includegraphics[width=\columnwidth]{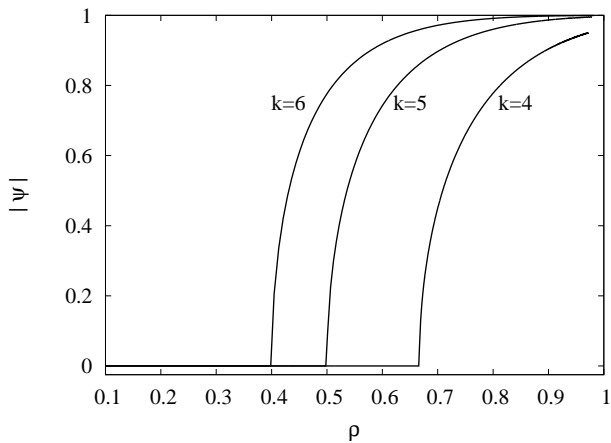}
\caption{The modulus of nematic order parameter $\psi$  as a function of the
density $\rho$ for different rod lengths $k$. The data are for $q=4$. 
} 
\label{fig03}
\end{figure}

Then at this point $f^* = g^* = x^*/k$, and $\rho_x = \rho_y = 1/(k-1)$. 
Correspondingly, we have $\rho_c = 2/(k-1)$. The corresponding value of 
critical activity, from Eq.~(\ref{eq:rhoq4}), is
\begin{equation}
z_c=\frac{(k-1)^{2 k -2}}{[k(k-3)]^k}, \quad q=4.\label{eq:zc}
\end{equation}
We note that the value of $z_c $ is finite only for $k \geq 4$.
In Fig.~\ref{fig03}, we show the variation of the modulus of the order
parameter $|\psi|$ with density $\rho$ for different rod lengths $k$.
$|\psi|$ is non--zero for densities larger than the critical density.

\section{\label{qhigherfour} Lattices with coordination number $q \geq 6$}

The analysis of $q \geq 6$ is very similar.  It is easy to check that for 
general even $q$, the fixed point solution which is independent of the
layer index $j$, satisfies the self-consistent equations 
\begin{equation}
k z ( 1 - \rho)^k  =  \rho_i  \left[ 1 - \frac{k-1}{k}
\rho_i \right]^{k-1},~i=1,\ldots,\frac{q}{2},
\label{eq:rhoqg4}
\end{equation}
where $i$ labels the $q/2$ different bond types, and $\rho=\sum_i
\rho_i$ is the total density of sites that are part of $k$-mers. 
The entropy per site generalises to
\begin{eqnarray}
s&=& \sum_{i=1}^{q/2} \left(1-\frac{k-1}{k}\rho_i\right)\ln
\left(1-\frac{k-1}{k}\rho_i\right)
\nonumber \\
&& - (1-\rho)\ln (1-\rho)
- \sum_{i=1}^{q/2} \frac{\rho_i}{k}\ln
\frac{\rho_i}{k}.
\label{entrop6}
\end{eqnarray}

\begin{figure}
\includegraphics[width=\columnwidth]{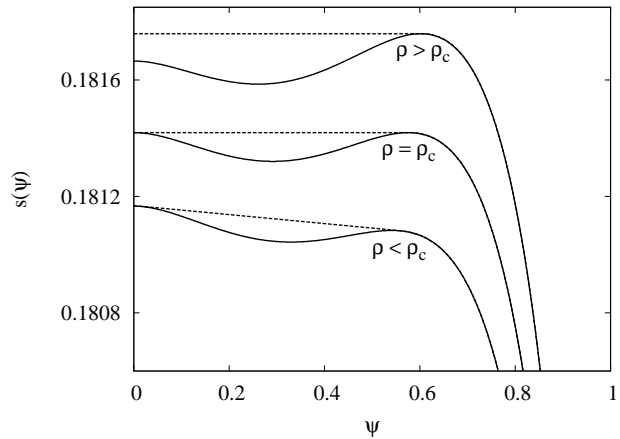}
\caption{Entropy as a as a function of the order parameter $\psi$ for
different densities. The data are for $q=6$, $k=8$, when the
isotropic--nematic transition is first order. The different curves
correspond to (a) $\rho =0.3890$, (b) $\rho= 0.3910 \approx \rho_c$, (c)
$\rho=0.3930$. The dotted line shows the convex envelope.
} 
\label{fig:q6k8entropy}
\end{figure}

The low density phase is isotropic, with 
$\rho_i$ same for different $i$. We define the order parameter to be
$\psi= (\rho_1 - \rho_2)/\rho$, with $\rho_2=\ldots=\rho_{q/2}$.
But now the entropy function has no symmetry under $\psi \rightarrow - 
\psi$. Then, the expansion of $s(\rho_i)$ in 
powers of $\psi$ contains cubic terms. Fig.~\ref{fig:q6k8entropy} shows the 
behaviour of entropy as a function of $\psi$ for different densities $\rho$
when $q=6$.  For small 
$\rho$, there is a single maximum at $\psi$ equal to zero. For 
larger $\rho$, a second local maximum at a non-zero $\psi$ appears, 
and at some value of $\rho$ this becomes of equal height. Then the order 
parameter jumps discontinuously, as the density is increased.

\begin{figure}
\includegraphics[width=\columnwidth]{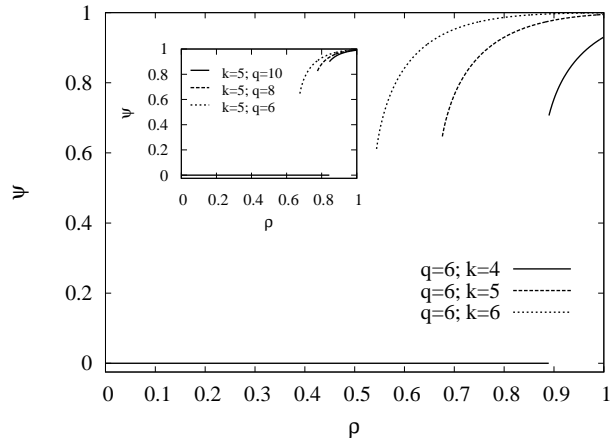}
\caption{\label{qhigherorder_rho}
The order parameter as a function of density $\rho$ for $q=6$ and varying $k$.
There is a first order transition at a critical value $\rho_c$ 
which decreases with $k$. Inset: The order parameter $\psi$ as a function of 
$\rho$ for $k=5$ and varying $q$. $\rho_c$ increases with $q$.
}
\end{figure}

In Fig.~\ref{qhigherorder_rho}, we show the variation of the order
parameter $\psi$  with density $\rho$ for $q=6$ and different values of $k$, 
and for $k=5$ and different values of $q$. The first
order transition is clearly seen, with the critical activity increasing with
$q$ and decreasing with $k$.

\begin{figure}
\includegraphics[width=\columnwidth]{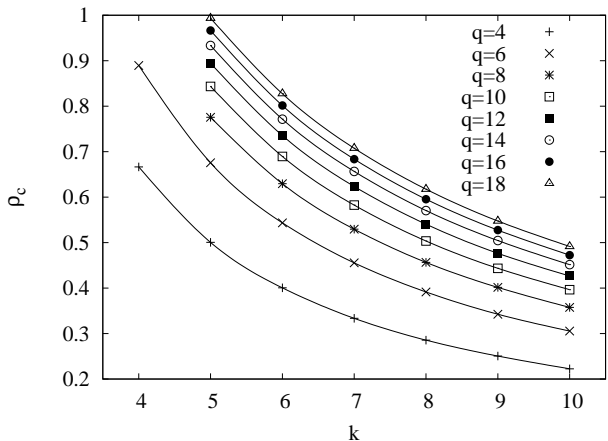}
\caption{\label{criticalrho}
The critical density $\rho_c$ of the isotropic--nematic phase
transition as a function of $k$ for different values of the
coordination number $q$. For $q=4$ the transition is continuous.} 
\end{figure}

For coordination number greater than four, it is not possible to
determine the critical density exactly. The critical densities obtained
by numerically comparing the entropy of the isotropic and nematic
phases are summarised in Fig.~\ref{criticalrho}. For a fixed value of
$q$, $\rho_c \sim \ln(q)/k$ for large $k$.

\begin{figure}
\includegraphics[width=\columnwidth]{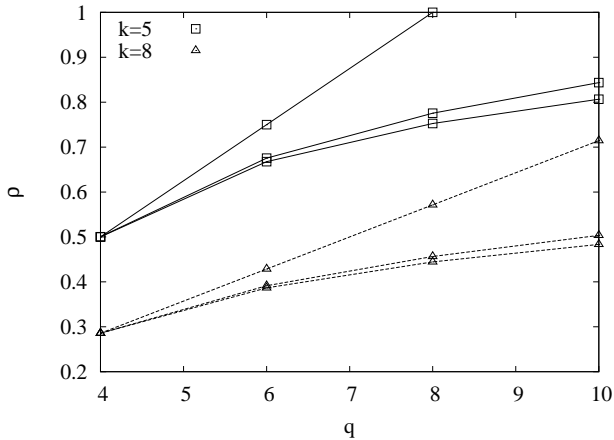}
\caption{\label{spinodals}
Coexistence and limits of stability values of density of the isotropic and
nematic phases as functions of $q$ for two different rod lengths. For
given values of $q$ and $k$, the lowest point is
the is the density at which $s(\psi)$ develops a  second local maximum, 
the intermediate point is the critical density $\rho_c$, and the highest point
is the density at which $s(\psi)$ is no longer a local maximum at
$\psi=0$.  The lines are guides to the eye.}
\end{figure}

As expected in discontinuous transitions, in a range of values for the
densities around the critical density $\rho_c$, entropy has a local maximum for
both the isotropic ($\psi=0$) and nematic phases. 
In Fig.~\ref{spinodals}, the values of densities at which these local
maxima appear and disappear, along with the critical density are shown
for $k=5$ and  $k=8$ for different values of $q$. 
Only for $q=4$, where the transitions are continuous, all the
densities coincide.

It is possible to determine the spinodal density $\rho_s$, the density at 
which the entropy $s(\psi)$ no longer has a maximum at $\psi=0$.
As in the analysis of $q=4$, the spinodal is still 
given by the condition $x^* = 1/(k-1)$. But now that there are $q/2$ 
distinct directions, the density at the spinodal point is $q/[2(k-1)]$. If 
$q < 2(k-1)$, the spinodal density is less than one, and entropy will
have a maximum at the nematic fixed point.
Thus, there is always a phase transition for $q< 2(k-1)$.

We now address the question of whether for coordination number $q$, 
there is a minimum rod length $k_{min}$ below which 
the isotropic--nematic transition is absent. From the analysis of the
spinodal,
\be
k_{min} \leq \frac{q+4}{2}.
\ee
The precise value of $k_{min}$ is computed numerically and the results
are shown in Table~\ref{table1}. We find that $k_{min} \sim \ln(q)$.
\begin{table}
\caption{\label{table1} The minimum rod length $k_{min}$ required for an
isotropic--nematic transition as a function of coordination number $q$.}
\begin{ruledtabular}
\begin{tabular}{l r| l r}
q&$k_{min}$ & q&$k_{min}$\\\hline
$q\in [4,6]$ & 4 & $q \in[656,1612]$ & 10 \\
$q \in [8,18]$ & 5 & $q \in[1614,3994] $ & 11\\
$q \in [20,44]$ & 6 & $q \in[3996, 9968] $ & 12\\
$q \in [46,110]$ & 7  &$q \in[9970, 25028] $ & 13 \\
$q \in [112,266]$ & 8  &$q \in[25030, 63188] $ & 14 \\
$q \in [268,654]$ & 9 & &  
\end{tabular}
\end{ruledtabular}
\end{table}

\section{Discussion}
\label{discussion}

We are not able to address the nature of the second transition in this 
study, as on the RLTL lattice, there is none. However, on this lattice, 
the limit of fully packed lattice is quite interesting.  We find that in 
this limit, the system has long-range nematic order, but the ordering is 
not complete, and there are small islands of ``wrongly'' oriented 
$k$-mers in a sea of aligned $k$-mers.  The small concentration of these 
wrongly oriented rods is entropically stabilised.

The order parameter $\psi$ is easy to determine for $k=4$, when
\begin{equation}
\psi=
\begin{cases}
\frac{5\sqrt{3}}{9} \approx 0.96225, & \text{k=4, q=4,} \\
\frac{3(\sqrt{6}-4)}{5} \approx 0.93031, & \text{k=4, q=6,} \\
\frac{13}{15} \approx 0.86667, & \text{k=4, q=8.}
\end{cases}
\label{dri} 
\end{equation}
The limiting value of $\psi$ grows monotonically
with $k$, being equal to 1 in the limit $k \to \infty$. 
It is easy to see that the
nematic order parameter has the asymptotic behaviour
\begin{equation}
1-\psi \approx \frac{q}{2 k^{k-1}},\quad k\rightarrow
\infty.
\label{eq:expansion}
\end{equation}

We can also look for a periodic solution of period $k$, where  
$x_{m+k}= x_m, y_{m+k} = y_m$.
In this case, the $2k$ independent parameters $f^*_s, g^*_s$, with $s
= 0$ to $k-1$ satisfy the equations 
\begin{eqnarray}
{f^*_j} ( 1 - \rho_1 + f^*_j) &=& z_1  ( 1 - \rho_1 - \rho_2)^{k} /A,
\\ \nonumber
g^*_j ( 1 - \rho_2 + g^*_j)& =& z_2  ( 1 -\rho_1 - \rho_2)^{k} /B,
\label{eq:periodic}
\end{eqnarray}
where
\begin{equation}
\rho_1 = \sum_{s=0}^{k-1} f^*_{s} ; \rho_2 =\sum_{s=0}^{k-1} g^*_{s},
\end{equation}
and
\begin{equation}
A =  \prod_{s=0}^{k-1} \left[ 1 - \rho_1 + f^*_{s}\right]; B=
\prod_{s=0}^{k-1} \left[ 1 - \rho_y + g^*_{s}\right]. 
\end{equation}

In Eq.~(\ref{eq:periodic}), the left hand side is a function of the 
form $( x + c x^2)$, with $c$ positive, and the right hand side is 
independent of $j$. Hence, the only positive real solution of this 
equation is of the form $f^*_j $ independent of $j$, and we do not get a 
non-trivial periodic solution. Note that a periodic solution would 
correspond to smectic-like layered ordering, and our solution rules it 
out.

We can easily extend our treatment to semi-flexible rods, where all the 
rods are aligned in the direction of increasing layer number, but a  
$k$-mer lying on an X-bond between layers $j$ and $(j=1)$ can bend, and 
lie on a Y-bond between layers $(j+1)$ and $(j+2)$, with some energy 
cost.  Solutions for flexible and semi-flexible polymers
on Bethe and Husimi lattices can be found in 
Refs.~\cite{stilck1990,stilck1993}.

There are other models like the ANNNI model \cite{selke1988}, where exact 
solution in dimensions greater than one is not possible, and the 
equilibrium state shows spatial structure.  These are usually discussed 
in the spatially varying mean-field approximation.  We feel that studying 
such model on RLTL lattice can take into account the short-range 
correlations in these systems better.

\begin{acknowledgments}
The work of DD and RR was partially supported by the Department of
Science and Technology, Government of India under the project
DST/INT/Brazil/RPO-40/2007, and that of JFS by
CNPq under the project 490843/2007-7. We thank Ronald Dickman for a careful
reading of the manuscript.
\end{acknowledgments}

%Merlin.mbs v4.21 2009-07-09.
%

%\bibliography{ref}

\end{document}